# Variation in Microbial Growth under Hypergravity

Conrad Cabral[1], Chintamani Pai[2], Kashmira Prasade[3], Smruti Deoghare[1], Urooz Kazi[4], Sonalia Fernandes[4]

*Department of Zoology, St. Xavier's College, Mahapalika Marg, Fort, Mumbai – 400001[1]*
*Department of Physics, University of Mumbai, Kalina Campus, Santacruz, Mumbai - 4000098[2]*
*Department of Microbiology, St. Xavier's College, Mahapalika Marg, Fort, Mumbai – 400001[3]*
*Department of Life Sciences, St. Xavier's College, Mahapalika Marg, Fort, Mumbai – 400001[4]*

*Corresponding Author: conrad.cabral@xaviers.edu
Tel. No.: +91-022-22620661
Fax No.: +91-022226594

**Abstract**

We report bacterial growth under hypergravitational stress. Cultures of *E. coli* and *B. subtilis* were subjected to the gravitational stress (38g) and their growth curves were measured using UV-VIS spectrophotometer. Experiments were also carried out to investigate nutrient consumption under hypergravitational conditions. Our results show considerable difference between samples subjected to hypergravity and normal conditions. This study has importance to understand bacterial response to external stress factors like gravity and changes in bacterial system in order to adapt with stress conditions for its survival.

**Introduction**

Studies have been carried out to understand microbial cell viability in different conditions like mechanical stress, ultrasonic vibrations [1], shearing forces, abrasive treatment [2], high magnetic field [3] and effect of high pressure [4]. Apart from these, specific studies related to microgravity environment have also been well documented. Cell response to microgravity suggests that it enhances growth rates in bacterial cultures [5-7]. Most microgravity studies have been carried out either in space shuttles or on ground based space flight simulators.

The opposite phenomenon called hypergravity has been an area of interest in recent times. Eukaryotic systems exhibit a very poor adaptation to hypergravity and have reduced cellular functions [8]. Different speculations have been made about cell regulatory factors like apoptosis, cell differentiation, growth and cell viability when exposed to gravitational stress [9-10]. Many studies report cell viability in hypergravitational conditions in a variety of g scales ranging from 4 g to $4x10^5$ g. Feasibility of such conditions in outer space environment where, microbes may be present in asteroids or meteorites and exposed to such wide range of g forces must be considered [10-12]. There have been efforts to understand molecular level controls for the above mentioned cell alterations observed during hypergravity. The following paper concentrates on studying the bacterial cell response under hypergravity (38g) by measuring bacterial growth curve and nutrient consumption rates.

## Materials and methodology

### Growth Curve Experiments

*E.coli* (K12) a gram negative rod is found in normal gut flora of warm blooded organisms. B. subtilis is a gram positive rod found mainly in soil. The bacteria were selected because of their opposite gram nature and the fact that *B.subtilis* is a sporulating bacterium while *E.coli* is not. Overnight grown cultures at 37$^{o}$C were inoculated in sterile Luria broth and incubated at 37$^{o}$C in a waterbath incubator Shaker (Trishul) at 150 rpm. Aliquots were taken at stipulated time points under sterile conditions. These aliquots were assayed for optical density (OD) at 600 nm in a UV-Vis spectrophotometer (Shimadzu UV-VS 1700). This procedure was followed for *E.coli* and *B.subtilis* control samples i.e. 1 x g.

For simulating the hypergravity condition a centrifuge (REMI-RMI2C) was used. The radius of the rotor was 6 cm to the point of the tubes and the angle of the tubes with respect to the horizontal were 30$^{o}$. The cultures were grown in sterile microcentrifuge tubes (MCTs) (1.5ml) (Axygen). The initial inoculum concentration was the same in all tubes (~$10^9$ per ml) that were subject to hypergravity and control conditions. The experimental tubes were spun at 750 rpm thus inducing a gravitational acceleration of approximately 38 x g. Precautions were taken to avoid cell pelleting in the MCTs. At stipulated time intervals, each MCT was taken out from the centrifuge and assayed in similar fashion as the control samples. All experiments were done in triplicates.

### Measuring Cell viability

Saline suspensions of the culture was made and subjected to the conditions of normal growth and hypergravity to rule out cell death occurring during hypergravity due to cell to cell collision or frictional forces arising due to centrifugation and viscosity of liquid,. The suspensions were then assayed at specific time intervals during bacterial growth curves to check cell viability. Cell viability was assayed using the colony forming unit (CFU) method for enumerating bacterial cells.

### Nutrient Consumption Assay

The nutrient consumption assay was done to check whether the reduced growth rate is a direct result of the reduction in nutrient intake of cells. M9 minimal media supplemented with 1% Glucose was used as a nutrient source. The inoculum load was the same as in the growth curve experiments. The glucose concentrations was assayed using the DNSA method of Glucose estimation. The remnant glucose gives a direct estimation of glucose consumption at different time intervals of bacterial growth in controlled and experimental (hypergravity) conditions. DNSA was assayed UV-VIS (Shimadzu UV-VS 1700) at 540 nm.

**Results and discussions**

*E.coli* and *B. subtilis* showed a remarkable difference in growth marked by OD, between the control and hypergravity treated cells (fig. 1 & fig. 2). When compared with the control, the treated cells show a reduction in growth starting from the time the cells enter the logarithmic phase of growth. The lag phase shows no noticeable difference between the two cultures as this is considered the phase of acclimatization. The difference between the growth rates of both the control and treated cells was observed to vary from a minimum of 24% to a maximum of 55% in *E.coli* and from 12% to 61% in *B.subtilis.* Both cultures (controls and experimental) entered the log and stationary phases almost simultaneously. No previous studies have reported growth curve analysis of the cells especially under hypergravity.

Cell viability was measured to investigate reasons behind the growth patterns exhibited by the cultures subjected to hypergravity. It is interesting to find, that though both cultures show noticeable difference in growth rates between control and hypergravity cells, there was no significant difference between the cell viability counts for both. This indicates that the reduced growth rate is due to changes occurring in the cell in order to adapt to hypergravitational stress rather than cell death occurring due to internal local factor like collisions and friction among bacterial cells. This supports a previous study based on cell viability under similar conditions [10].

DNSA was done to understand the factors influencing cell growth when subjected to hypergravity (fig. 3). The result shows that the glucose consumption for both (control and hypergravity) cultures were different. The control cells reached a state of complete nutrient depletion in 6 hrs well before the experimental cells which took 10 hours to reach the same state. This study shows that metabolic activities in the cell are affected due to hypergravity. This may give an insight to understand system biology of bacterial cell and its dynamics during stress signal transduction.

A growth curve experiment was done using M9 minimal media and growth rates were obtained for both hypergravity and control cultures. The growth rate obtained was then superimposed with the glucose consumption graph for Control and Hypergravity cultures (fig. 4). The graph shows that though there is a reduction in cell density in hypergravity treated cells, both cultures show maximum rate of growth at the same time approximately 240 minutes. This is in accordance with our above growth curve data in which, it is seen that though cell density is reduced considerably in hypergravity treated cells, they do enter log and stationary phases at the same time points as the control cells.

The work done clearly shows that conditions like hypergravity affect cellular growth and nutrient uptake of bacterial cell. Even though the cells were subjected to 38g they showed a typical growth curve pattern with reduced growth rates. It is interesting to note that the cell duplication time increased 3 fold the normal for *E.coli* from 25 min. for a control culture to 72 min. in the hypergravity state. This can lead to understanding the cell adaptation process in the cells when subjected to hypergravity. Such experiments can reveal bacterial growth in outer space rocks undergoing high accelerations, collisions and their response to external stress factors.

## Acknowledgement

The authors thank Principal, St. Xavier's college for the support given to carry out the experiments in the Central Instrumentation Facility (CIF). We are grateful to Dept. of Zoology, Dept. of Microbiology, Dept. of Botany St. Xavier's College for their constant support. We are also grateful to Mr. Melwyn for the maintenance of our set-up.

**Figure Legends:**

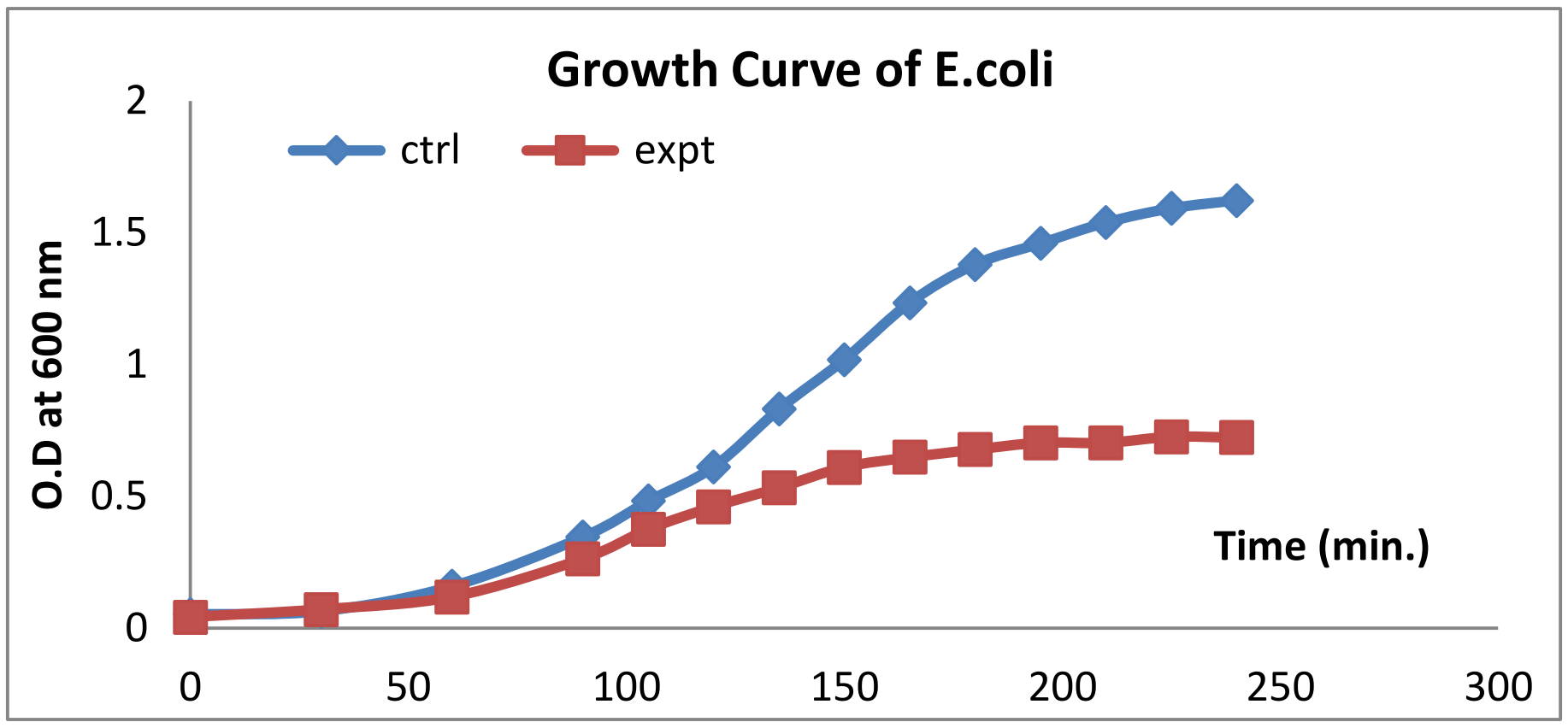

**Fig. 1:** Growth curve of *E.coli* under control and hypergravity conditions.

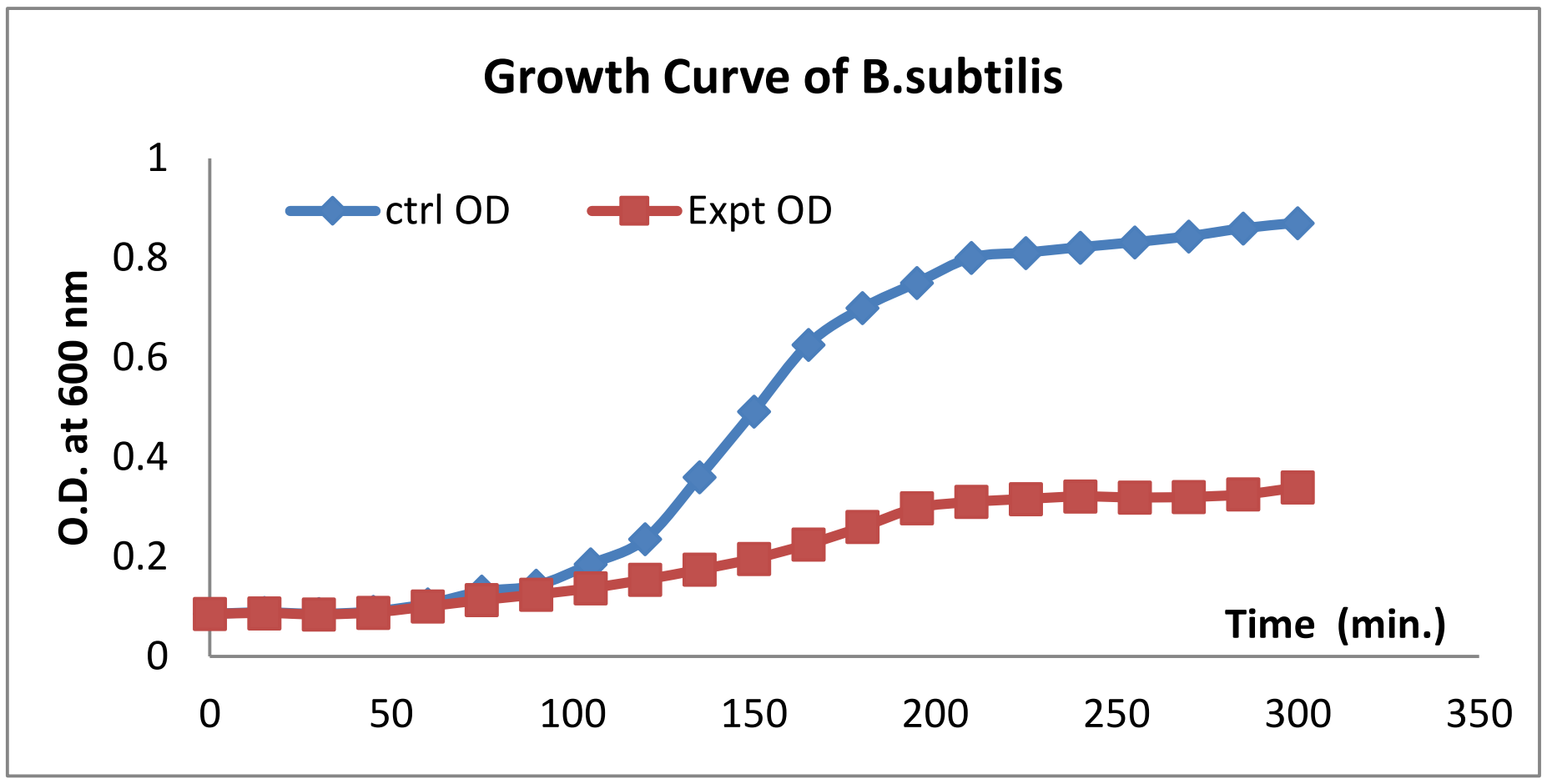

**Fig. 2:** Growth curve of *B.subtilis* under control and hypergravity conditions.

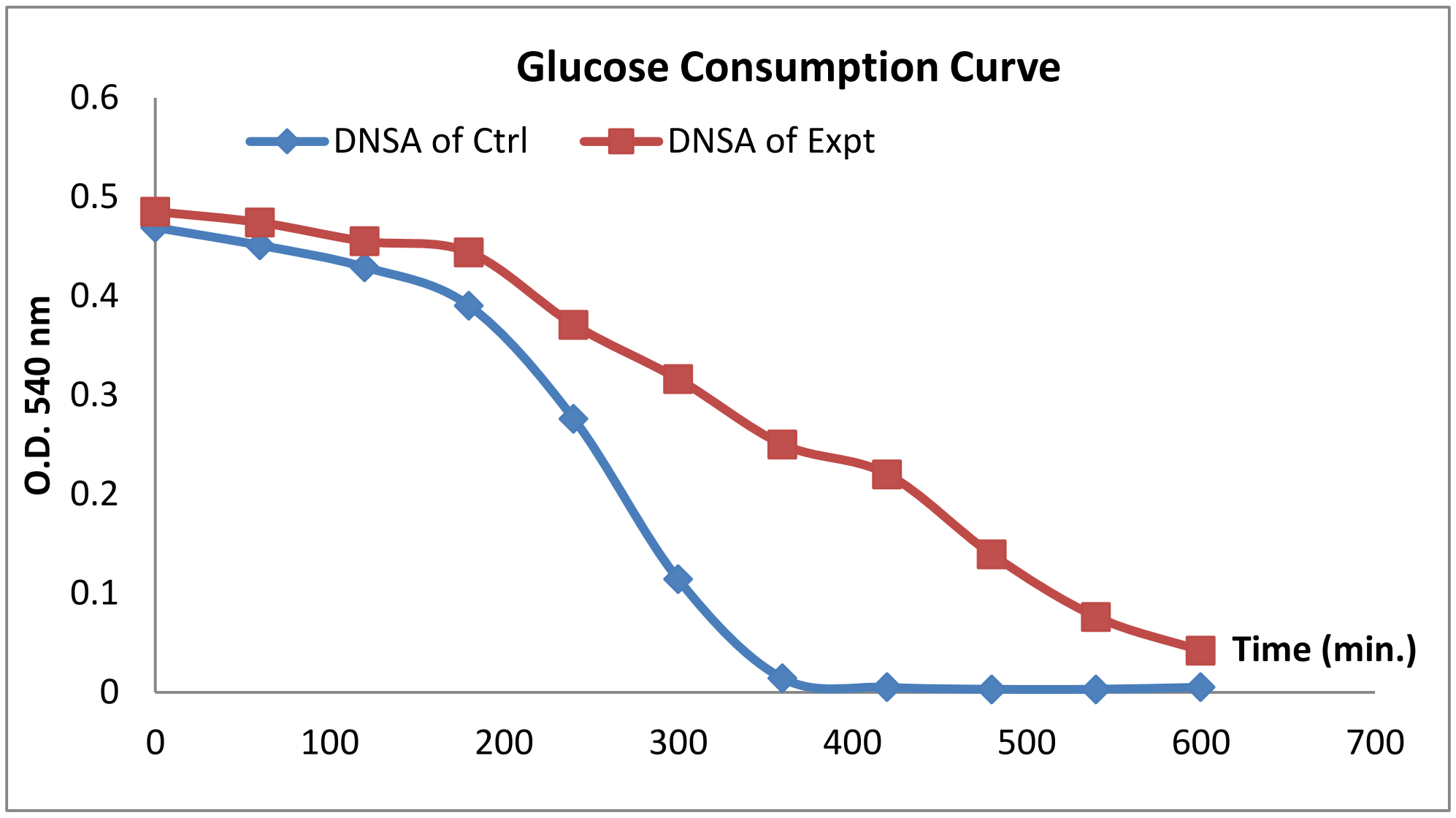

**Fig. 3:** Glucose consumption assay (*E.coli*) for control and hypergravity treated cells.

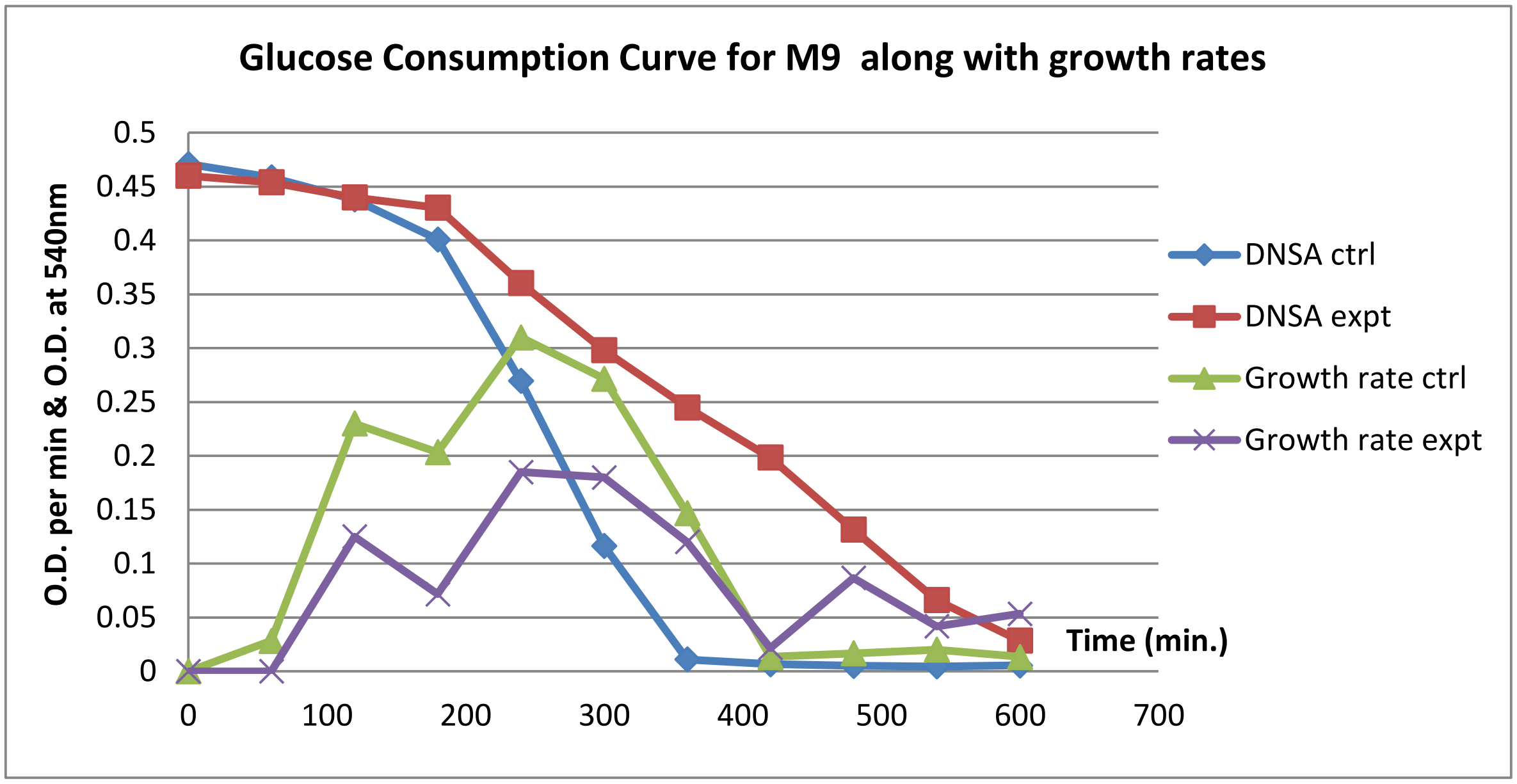

**Fig. 4:** Growth rates superimposed with Glucose consumption for hypergravity and control cells in M9 minimal media for *E.coli*.